# Design-Oriented Transient Stability Analysis of PLL-Synchronized Voltage-Source Converters

Heng Wu, *Student Member*, *IEEE*, and Xiongfei Wang, *Senior Member*, *IEEE*
Department of Energy Technology, Aalborg University, 9220 Aalborg East, Denmark

*Abstract*---Differing from synchronous generators, there are lack of physical laws governing the synchronization dynamics of voltage-source converters (VSCs). The widely used phase-locked loop (PLL) plays a critical role in maintaining the synchronism of current-controlled VSCs, whose dynamics are highly affected by the power exchange between VSCs and the grid. This paper presents a design-oriented analysis on the transient stability of PLL-synchronized VSCs, i.e., the synchronization stability of VSCs under large disturbances, by employing the phase portrait approach. Insights into the stabilizing effects of the first- and second-order PLLs are provided with the quantitative analysis. It is revealed that simply increasing the damping ratio of the second-order PLL may fail to stabilize VSCs during severe grid faults, while the first-order PLL can always guarantee the transient stability of VSCs when equilibrium operation points exist. An adaptive PLL that switches between the second-order and the first-order PLL during the fault-occurring/-clearing transient is proposed for preserving both the transient stability and the phase tracking accuracy. Time-domain simulations and experimental tests, considering both the grid fault and the fault recovery, are performed, and the obtained results validate the theoretical findings.

## I. INTRODUCTION

Voltage-source converters (VSCs) are commonly used with renewable energy resources, flexible power transmission systems and electrified transportation systems. The ever-increasing penetration of VSCs is radically changing the dynamic operations of power grids. Differing from synchronous generators, the dynamic behavior of the VSC is highly affected by its control algorithms. The instability phenomena resulted from the control dynamics of VSCs under different grid conditions are increasingly reported, ranging from the harmonic stability to the loss of synchronization (LOS), which severely challenge the security of electricity supply in the power grid with high penetration of renewables [1].

There have been increasing research efforts spent on tackling the stability challenges brought by VSC-grid interactions. The small-signal modeling and stability analysis of power-electronic-based power systems have been thoroughly discussed in recent years [2]. It is found that the phase-locked loop (PLL) brings a negative damping within its bandwidth, which tends to destabilize the VSC under the weak (i.e. low short-circuit ratio) grid condition [3]-[4], and the asymmetric $dq$-frame control dynamic of the synchronous reference frame (SRF)-PLL leads to frequency-coupled oscillations [5].

In contrast, only a few research works can be found on the transient stability (i.e. the synchronization stability) of VSCs under large grid disturbances. In a recent report from North American Electric Reliability Corporation (NERC), the LOS of the PLL under grid faults has been found as one cause of the trip of a 900 MW photovoltaic power plant in Southern California [6]. Hence, the analysis of the impact of the PLL on the transient stability of VSCs during large disturbances is urgently demanded. Although a wide variety of improved PLLs with adaptive parameter-tuning during transient disturbances have been developed [7]-[9], only the dynamic performance of the PLL itself is considered in those works. The VSC-grid interaction, i.e., the voltage at the point of common coupling (PCC) used for the grid synchronization will be affected by the VSC's injected current, due to the voltage drop across the grid impedance, was overlooked. This VSC-grid interaction is the key factor leading to the LOS of the PLL As will be demonstrated in this paper, those second-order adaptive PLLs proposed in [7]-[9] still has a risk of LOS under large grid disturbances.

Considering VSC-grid interactions, a large-signal nonlinear model of the SRF-PLL is reported in [10], which reveals that the LOS will be inevitable if the VSC does not have equilibrium points during grid faults. Moreover, when there are equilibrium points during grid faults, the transient stability of the VSC is also analyzed by using the equal-area criterion (EAC) in [11]. Although the EAC-based analysis is intuitive with a physical insight, the conclusion is only valid when the proportional gain ($K_p$) of the SRF-PLL is zero, which is not feasible in practice. A more accurate analysis characterizing the dynamics of the SRF-PLL during grid faults is provided in [12]-[15] by using the phase portrait approach.

A number of control methods have also been developed for avoiding the LOS of the PLL-synchronized VSCs. The simplest approach is to freeze the PLL during grid faults [16], which is also recommended in the NERC report [6]. Yet, an

This paper has been partially presented at IEEE International Power Electronics Conference (IPEC-ECCE Asia), Niigata, Japan, 2018, and IEEE Energy Conversion Congress and Exposition (ECCE), Portland, USA, 2018. This work is supported by the Aalborg University Strategic Talent Management Programme. (*Corresponding author: Xiongfei Wang.*)
H. Wu and X. Wang are with the Department of Energy Technology, Aalborg University, 9220 Aalborg, Denmark (e-mail: hew@et.aau.dk; xwa@et.aau.dk).

obvious disadvantage of this method is that the VSC has no synchronization units during grid faults, and thus it fails to detect the right grid phase angle. Consequently, the injected active and reactive current is out of control during that period, which violates the grid code [17].

Besides freezing the PLL, the transient stability of VSCs can also be enhanced by directly modifying the injected active and reactive current to the grid, including the zero current injection [18], the adaptive current injection based on the X/R ratio of the grid impedance [19], and the adaptive current injection based on the detected frequency of the PLL [20]-[21]. The zero current injection method cannot comply with the grid code, which requires the VSC to inject 1.p.u. reactive current during severe fault [17]. The adaptive current injection based on the X/R ratio requires the prior knowledge of the grid impedance, which is also impractical. In contrast, the adaptive current injection based on the output frequency of the PLL is more feasible, which, however, still may fail to inject 1 p.u. reactive current under the grid fault, when the grid impedance is not purely inductive [20]. To avoid modifying the injected current profile, the damping ratio of the SRF-PLL can be increased to enhance the transient stability of VSCs [12]-[15], [22]. Yet, those works have not quantified how large the damping ratio of the PLL is needed to stabilize VSCs, and thus the PLL design guideline is still missing.

This paper presents a design-oriented analysis of the transient stability impact of the PLL on current-controlled VSCs. Since the LOS of VSCs is inevitable when there is no equilibrium point during large disturbances [10], only the transient stability of VSCs with equilibrium points is considered. As an extension of the analysis in [12]-[13], this work first quantifies the critical damping ratio of the PLL under different operating conditions, i.e., different depths of grid voltage sags and different voltage drops across the line impedance. It is revealed that when only one equilibrium point exists during the fault, the SRF-PLL cannot stabilize VSCs no matter how large the damping ratio is adopted. Yet, this transient instability can be avoided if the SRF-PLL is reduced as a first-order PLL by freezing the integral controller. On the other hand, since the first-order PLL suffers from the steady-state phase error when the grid frequency has a steady-state drift from its nominal value [23], an adaptive PLL, which switches between the SRF-PLL and the first-order PLL based on the grid condition, is then developed to avoid the LOS of current-controlled VSCs. Moreover, instead of freezing the whole PLL, the adaptive PLL operates as the second-order PLL when the post-fault grid voltage comes to the steady state, which allows an accurate phase tracking during the grid fault and thus facilitates the fault recovery. Time-domain simulations and experimental tests validate the theoretical findings and the performance of the adaptive PLL.

## II. GRID-CONNECTED VOLTAGE SOURCE CONVERTERS

### A. System Description

Fig. 1 illustrates the simplified single-line diagram of a grid-connected three-phase VSC using the typical vector current control, where $L_f$ is the output filter of the converter, $Z_{line}$ represents the line impedance, $V_{gcp}$ and $\theta_{gcp}$ are the amplitude and the phase angle of the voltage at the grid connection point (GCP). The PCC voltage is measured for synchronizing the VSC with the grid by means of the PLL. $V_{PCC}$ and $\theta_{PCC}$ are the amplitude and the phase angle of the PCC voltage, and $\theta_{PLL}$ is the phase angle detected by the PLL. $\theta_{PLL}=\theta_{PCC}$ is expected at the steady state. $I_{dref}$ and $I_{qref}$ are current references for the active current and the reactive current, respectively. The current reference selection is switched to 1 during the normal operation, where $I_{dref}$ and $I_{qref}$ are determined by the dc-link voltage loop and the reactive power loop, respectively. During the grid faults, the current reference selection is switched to 2, where $I_{qref}$ is directly specified based on the requirement of the grid code, and $I_{dref}$ is changed based on $I_{qref}$ in order to avoid the overcurrent of the VSC [22]. The Proportional+ Integral (PI) controller is used for the current regulation in the $dq$-frame to guarantee a zero steady-state tracking error [24]. The outputs of the PI controller are transformed to the $abc$ frame and then fed into the pulse width modulation (PWM) block to generate drive signals of power switches.

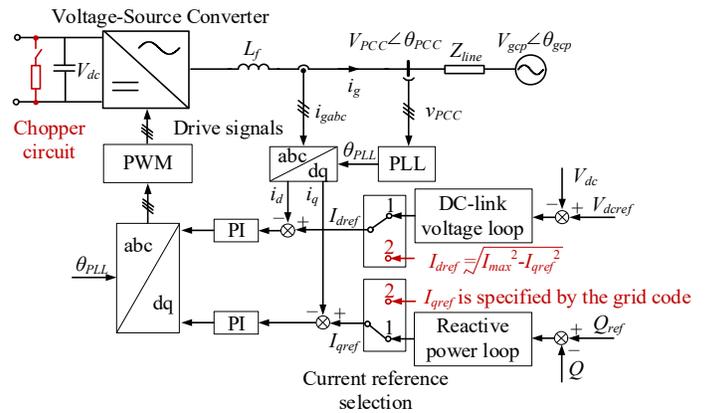

Fig. 1. A single-line diagram of a grid-connected VSC during the normal and fault ride-through operations. The current reference selection is switched to 1 during the normal operation and is switched to 2 during the fault ride through.

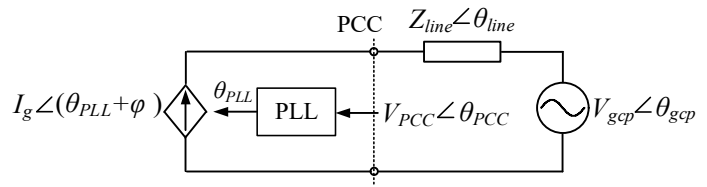

Fig. 2. The simplified converter-grid system for the transient stability analysis.

Based on the principle of model order reduction, the fast dynamics of the system can be neglected when analyzing the impact of the slow dynamics [25]. The bandwidth of the PLL is designed much lower than that of the inner current loop [26]. Thus, the inner current loop can be approximated as a unity gain in the transient stability study. As the dc-link voltage loop and the reactive power loop are de-activated during the grid fault, the synchronization stability of the VSC during the grid fault is dominated by the dynamics of the PLL. Consequently, the system diagram shown in Fig. 1 can be simplified as a controlled current source with its phase angle regulated by the PLL, as shown in Fig. 2. $\varphi$ denotes the phase difference between the PCC voltage $v_{PCC}$ and the grid current $i_g$, which is also called the power factor angle. $I_g$ is the amplitude of the grid current. $\theta_{line}$ represents the line impedance angle. It is worth mentioning that this simplified representation of grid-connected VSCs has been proven as adequate in [10]-[16], [18]-[22] for analyzing the transient stability impact of the PLL.

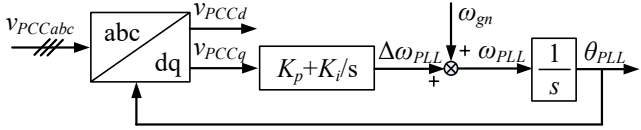

Fig. 3. Block diagram of the SRF-PLL.

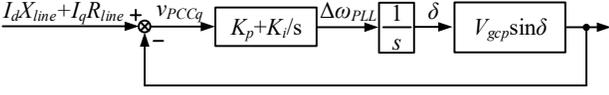

Fig. 4. The equivalent diagram of the SRF-PLL considering the effect of the line impedance.

### B. Mathematical Model of the PLL Considering Line Impedance Effect

Fig. 3 illustrates the block diagram of the commonly used SRF-PLL [23]. The three-phase PCC voltages are sampled and transformed to the *dq* frame. The *q*-axis voltage is regulated by a PI controller for tracking the grid phase [23].

Based on Fig. 3, the dynamic equation of the PLL can be expressed as

$$\theta_{PLL} = \int \left[ \omega_{gn} + \left( K_p + K_i \int \right) v_{PCCq} \right]. \quad (1)$$

where $\omega_{gn}$ is the nominal grid frequency. $K_p$ and $K_i$ are the proportional and integral gain of the PI regulator, respectively. $v_{PCCq}$ represents the *q*-axis component of the PCC voltage.

Based on Fig. 2, $v_{PCCq}$ can be calculated as

$$v_{PCCq} = v_{zq} + v_{gcpq} \quad (2)$$

where $v_{zq}$ and $v_{gcpq}$ denote the *q*-axis component of the voltage across the line-impedance and the GCP voltage, respectively. The line impedance can be given by $Z_{line}=R_{line}+jX_{line}$, and $v_{zq}$ and $v_{gcpq}$ can then be derived as

$$v_{zq} = I_d X_{line} + I_q R_{line}. \quad (3)$$

$$v_{gcpq} = V_{gcp} \sin\left(\theta_{gcp} - \theta_{PLL}\right). \quad (4)$$

Defining the angle difference between $\theta_{PLL}$ and $\theta_{gcp}$ as $\delta$, i.e.,

$$\delta = \theta_{PLL} - \theta_{gcp}. \quad (5)$$

Substituting (2)-(5) into (1) and considering the integral relationship that $\theta_{gcp}=\int\omega_{gn}dt$, which yields

$$\delta = \int \left( K_p + K_i \int \right)\left( I_d X_{line} + I_q R_{line} - V_{gcp} \sin\delta \right). \quad (6)$$

Thus, the second-order phase-swing behavior of the PLL that considers the VSC-grid interaction can be characterized by (6). On the basis of this, the equivalent diagram of the SRF-PLL can be drawn, as shown in Fig. 4.

### C. Equilibrium Points

The stable operation of the system requires the existence of equilibrium points, where $v_{PCCq}=0$, leading to

$$I_d X_{line} + I_q R_{line} = V_{gcp} \sin\delta \quad (7)$$

The existence of the solution of (7) requires

$$\left| I_d X_{line} + I_q R_{line} \right| \leq V_{gcp} \quad (8)$$

It is known from (8) that the existence of the equilibrium points is affected by the injected active current and reactive current, the grid impedance and the grid voltage magnitude during the fault. Eq. (8) is always satisfied if $I_d X_{line}+I_q R_{line}=0$, which can be realized by choosing $I_d=I_q=0$ [18] or $I_d/I_q=-R_{line}/X_{line}$ [19]. However, these solutions violate the grid code [17]. Considering the specific amount of the injected active current and reactive current required by the grid code, the loss of equilibrium points is more likely to happen during severe fault (small $V_{gcp}$) under the weak grid condition (large $X_{line}$ and $R_{line}$). The LOS will be inevitable if there is no equilibrium point during grid faults [10], and it may also take place even if the equilibrium point exists [12], which will be detailed in the next section.

## III. DESIGN-ORIENTED TRANSIENT STABILITY ANALYSIS

In this section, the LOS mechanism of VSCs during grid faults is elaborated. Depending on the number of equilibrium points, two operating scenarios of the VSC are considered, i.e., the VSC with two equilibrium points and the VSC with single equilibrium point during the grid fault.

Fig. 5 illustrates the voltage-angle curves of the VSC before and after the fault, which are drawn based on (6). The VSC is usually controlled with the unity power factor in the steady-state, i.e. $I_d=I_{max}$, $I_q=0$, where $I_{max}$ denotes the rated current of the VSC. Therefore, $v_{zq}$ can be simplified as $I_d X_{line}+I_q R_{line} = I_{max}X_{line}$. The dashed line in Fig. 5 illustrates the curve of $V_{gcp}\sin\delta$ before the grid fault, where the system is initially operated at the equilibrium point *a*, where $I_{max}X_{line} = V_{gcp}\sin\delta_0$.

Once the fault occurs, the GCP voltage magnitude drops to $V_{gcpfault}$ and the curve of $V_{gcpfault}\sin\delta$ is shifted as the solid line in Fig. 5. According to the grid code [17], the VSC needs to provide 2% reactive current per percent of the voltage drop, and a full reactive current is thus required when the GCP voltage is below half of the nominal value, i.e. $I_d = 0$, $I_q = -I_{max}$. As a consequence, $v_{zq}$ is changed as $I_d X_{line} + I_q R_{line} = -I_{max} R_{line}$. It is obvious that the VSC will lose the synchronism with the grid if $-I_{max} R_{line} < -(V_{gcpfault} \sin\delta)_{max} = -V_{gcpfault}$, since there is lack of equilibrium points [10].

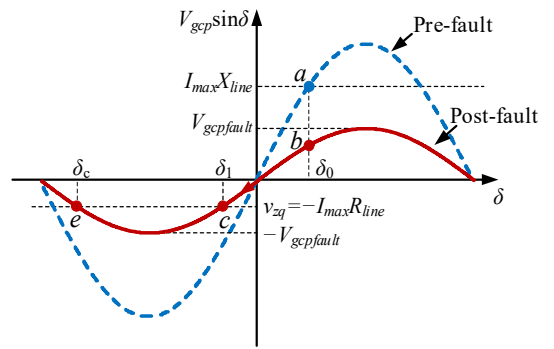

Fig. 5. Voltage-angle curves of the grid-connected VSC when $-I_{max}R_{line}>-V_{gcpfault}$.

On the other hand, the LOS can also arise even if there are equilibrium points during the fault [12]. From the solid line in Fig. 5, it is clear that the system has two equilibrium points when $-I_{max} R_{line} > -V_{gcpfault}$, which are, similar to the power-angle curve of the synchronous generator, denoted as the stable equilibrium point (SEP) $c$ and the unstable equilibrium point (UEP) $e$ [27]. Yet, when $-I_{max} R_{line} = -V_{gcpfault}$, the system only has single equilibrium point.

### A. LOS Mechanism of VSCs with Two Equilibrium Points during Faults

As shown by the solid line in Fig. 5, the operating point of the system moves from the point $a$ to the point $b$ at the fault-occurring instant. Since $-I_{max} R_{line} < V_{gcpfault} \sin\delta_0$ (i.e., $v_{PCCq} < 0$) at the point $b$, the output frequency of the PLL starts to decrease, which leads to a decrease of $\delta$. The frequency continues to decrease until it reaches the SEP $c$, where $-I_{max} R_{line} = V_{gcpfault} \sin\delta_1$, as shown in Fig. 5. Since the output frequency of the PLL is below the grid frequency at the point $c$, $\delta$ continues to decrease, yet the output frequency of the PLL begins to increase after the point $c$ due to $-I_{max} R_{line} > V_{gcpfault} \sin\delta$ (i.e., $v_{PCCq} > 0$). Consequently, two possible operating scenarios can take place:

1) The output frequency of the PLL recovers to the grid frequency before the UEP $e$. As $-I_{max} R_{line} > V_{gcpfault} \sin\delta$ (i.e., $v_{PCCq} > 0$) still holds before the UEP $e$, the output frequency of the PLL further increases, which results in an increase of $\delta$. Thus, the operating point retraces the $V_{gcpfault} \sin\delta$ curve and finally reaches the SEP $c$ after several cycles of oscillation, and the system is stable.

2) The output frequency of the PLL is still below the grid frequency at the UEP $e$, Then, the output frequency turns to decrease again after the point $e$, due to $-I_{max} R_{line} < V_{gcpfault} \sin\delta$ (i.e., $v_{PCCq} < 0$), and $\delta$ keeps decreasing. The system eventually loses the synchronism with the grid.

### B. Parametric Effect of the PLL on Transient Stability of VSCs with Two Equilibrium Points during Faults

Following the analysis of the LOS mechanism of VSCs with two equilibrium points during faults, the parametric effect of the PLL is analyzed in this part. As a second-order dynamic system, the PLL are generally characterized by two important parameters, i.e. the damping ratio ($\zeta$) and the setting time ($t_s$) [23], which can be expressed by the controller parameters of the PLL as [24]:

$$\zeta = \frac{K_p}{2}\sqrt{\frac{V_{gn}}{K_i}}. \qquad (9)$$

$$t_s = \frac{9.2}{V_{gn} K_p}. \qquad (10)$$

where $V_{gn}$ is the nominal grid voltage. Applying the derivation on both sides of (6), and considering the relationship $X_{line} = (\omega_{gn} + \dot\delta) L_{line}$, which yields:

$$\ddot\delta = \frac{K_i}{1 - K_p I_d L_{line}}\left[I_d(\omega_{gn} + \dot\delta)L_{line} + I_q R_{line} - V_{gcp}\sin\delta\right] \\ - \frac{K_p V_{gcp}\cos\delta}{1 - K_p I_d L_{line}} \cdot \dot\delta. \qquad (11)$$

Then, substituting (9) and (10) into (11), the parametric effect of the PLL on the transient stability of the VSC during the grid fault can be evaluated. As for the second-order nonlinear dynamic system, the phase portrait approach provides an intuitive and design-oriented analysis [28]-[29].

#### 1) Influences of the settling time and damping ratio of the PLL

For illustrations, three phase portraits based on (11) are plotted in Fig. 6. The impacts of different settling times with the same damping ratio are evaluated in Fig. 6(a), while the effects of different damping ratios with the same settling time are analyzed in Fig. 6(b). Points $a$ and $c$ represent the SEPs of the system before and after the fault, respectively. It is clear that the system is stable when the phase portrait converges to the SEP $c$ after the fault, as shown by the solid and dashed lines in Fig. 6, and is unstable when the phase portrait is diverged, as shown by dashed-dotted lines in Fig. 6. Therefore, two important conclusions can be drawn:

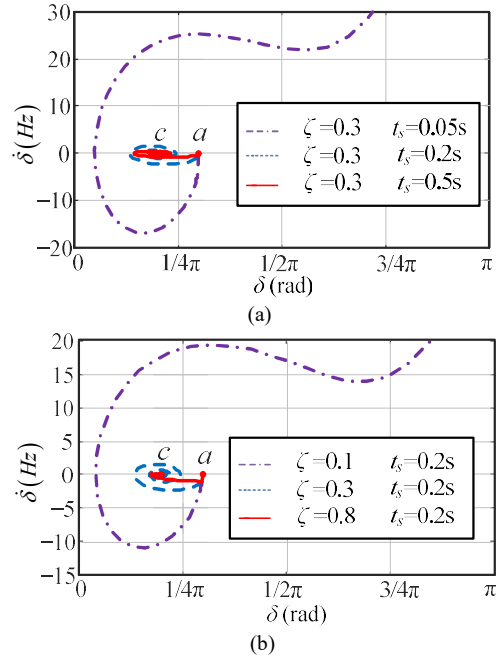

Fig. 6. Phase portraits of the PLL when $V_{gcp}$ drops from 1 p.u. to 0.6 p.u. (a) $\zeta=0.3$, $t_s=0.05$s (unstable), 0.2s (stable), 0.5s (stable). (b) $t_s=0.2$s, $\zeta=0.1$ (unstable), 0.3 (stable), 0.8 (stable).

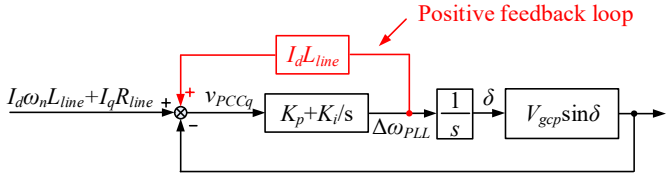

Fig. 7. Equivalent transformation of the block diagram of the SRF-PLL considering the effect of the frequency-dependent line reactance.

i) Reducing the settling time of the PLL jeopardizes the transient stability of the VSC, as shown in Fig. 6(a). This fact is resulted from the frequency-dependent nature of the line reactance, which inherently introduces a positive feedback loop. Considering the relationship $X_{line}=(\omega_n+\dot{\delta})L_{line}$, $v_{zq}$ in (3) can be rewritten as $v_{zq}=I_d\omega_nL_{line}+I_qR_{line}+I_d\dot{\delta}L_{line}$, and thus Fig. 4 can be transformed as Fig. 7, where a positive feedback loop is formed with the term $I_d\dot{\delta}L_{line}$. Since the smaller settling time implies a larger $K_p$, the loop gain of the positive feedback loop is consequently increased, and the transient stability is further deteriorated. Yet, if the line impedance is pure resistive ($L_{line}=0$) or there is no active current injection ($I_d=0$), the positive feedback loop shown in Fig. 7 will not exist and the settling time has no influence on the transient stability of the VSC, which has also been pointed out in [12].

ii) Increasing the damping ratio of the PLL enhances the transient stability of the VSC, as shown in Fig. 6(b). From the analysis based on Fig. 5, it is known that the LOS takes place when the PLL has a phase overshoot crossing the UEP $e$. This phase overshoot can be reduced by increasing the damping ratio of the PLL [23], and consequently the transient stability of the VSC is enhanced. This conclusion is similar to the system with synchronous generators (SGs), where the large damping term of the SG is also proven to be beneficial for its transient stability [27].

*2) Critical Damping Ratio of the PLL*

It is worth mentioning that the damping ratio of 0.707 is commonly chosen for the parameter tuning of the PLL [23]-[24], yet it is inappropriate when the transient stability of the VSC is concerned. Thus, the critical damping ratio that guarantees the transient stability of the system is quantified in the following.

Basically, the VSC will be stable after the large disturbance if $\delta$ is converged to the new equilibrium value, and will be unstable if $\delta$ is diverged. The trajectory of $\delta$ can be obtained by solving (11). As (11) is the second-order nonlinear differential equation, only numerical solution is possible [29]. The critical damping ratio is identified by an iterative calculation procedure, as shown in Fig. 8. The damping ratio is initialized with a small positive value (e.g. 0.1), and then increased with a fixed step size during every iteration until the solution (11) is converged. This procedure can be repeated to determine the critical damping ratios in different operating scenarios.

It is known from (3)-(6) that the dynamic of $\delta$ is affected not only by the post-fault grid voltage $V_{gcpfault}$, but also by the $q$- axis voltage drop across the line impedance $v_{zq}$. Therefore, the critical damping ratios with different $|v_{zq}|/|V_{gcpfault}|$ ratios are plotted in Fig. 9. It is clear that the higher $|v_{zq}|/|V_{gcpfault}|$ ratio, the larger critical damping ratio is required. Moreover, when $|v_{zq}|/|V_{gcpfault}|=1$, which corresponds to the case that the PLL has single equilibrium point, $\delta$ cannot be converged no matter how large the damping ratio is adopted. This phenomenon will be analytically explained in the next part of this section.

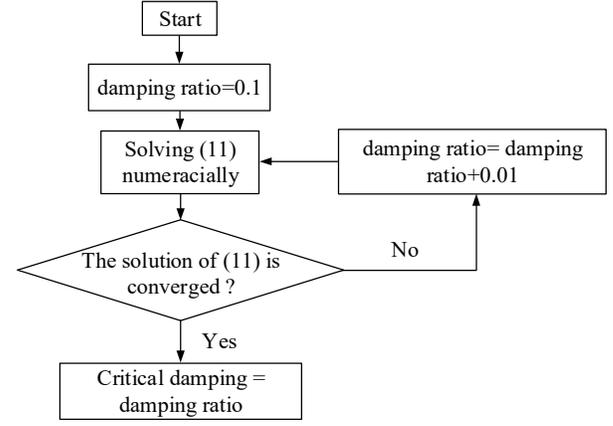

Fig. 8. Iterative calculation procedure for the critical damping ratio.

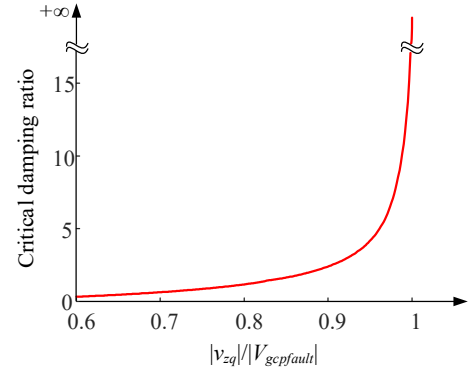

Fig. 9. Critical damping ratio with different $|v_{zq}|/|V_{gcpfault}|$ ratios.

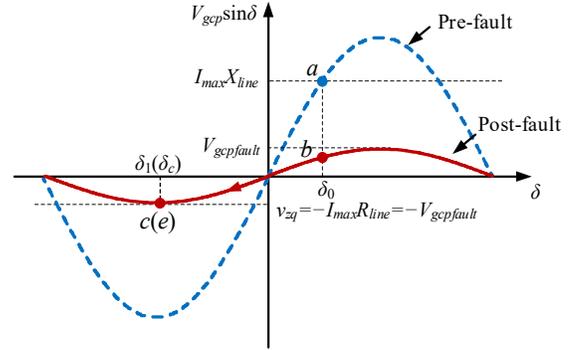

Fig. 10. Voltage-angle curves of grid-connected VSCs when $v_{zq}=-I_{max}R_{line}=-V_{gcpfault}$.

*C. Transient Stability of VSCs with Single Equilibrium Point during Faults*

Fig. 10 illustrates the voltage-angle curve of the VSC under the condition of $v_{zq}=-I_{max}R_{line}=-V_{gcpfault}$. In this case, the SEP $c$ and the UEP $e$ merge as a single point, and the output frequency of the PLL has to be recovered to the grid frequency at the point $c(e)$ in order to remain the synchronism with the grid, i.e., $\dot{\delta}=0$ at the point $c(e)$. Consequently, the dynamic response of $\delta$ must be overdamped for a stable operation. Any small phase overshoots in the dynamic response can make the system cross over the point $c(e)$, and eventually result in the LOS.

Applying the derivation on both sides of (6) yields

$$\dot{\delta} = \left(K_p + K_i \int\right)\left(I_d X_{line} + I_q R_{line} - V_{gcp}\sin\delta\right)$$
$$= \left(K_p + K_i \int\right)\left(-I_{max} R_{line} - V_{gcpfault}\sin\delta\right) = \dot{\delta}_1 + \dot{\delta}_2 \quad (12)$$

where

$$\dot{\delta}_1 = K_p\left(-I_{max} R_{line} - V_{gcpfault}\sin\delta\right) \quad (13)$$

$$\dot{\delta}_2 = K_i \int\left(-I_{max} R_{line} - V_{gcpfault}\sin\delta\right) \quad (14)$$

Since $-I_{max}R_{line} = V_{gcpfault}\sin\delta$ holds at the equilibrium point $c(e)$, it is known from (13) that $\dot{\delta}_1 = 0$ at the point $c(e)$. Yet, $-I_{max}R_{line} < V_{gcpfault}\sin\delta$ always holds during the dynamic process when the operating point moves from the point $b$ to the point $c(e)$ under grid faults, as shown in Fig. 10. This makes the integration of $(-I_{max}R_{line} - V_{gcpfault}\sin\delta)$, i.e., $\dot{\delta}_2$, smaller than zero at the point $c(e)$. Consequently, the condition $\dot{\delta} = \dot{\delta}_1 + \dot{\delta}_2 = \dot{\delta}_2 < 0$ always holds at the point $c(e)$ as long as $K_i > 0$. This fact implies that the SRF-PLL will cross over the point $c(e)$, and thus the LOS is inevitable for the VSC when there is only one equilibrium point during the fault, no matter how large the damping ratio is adopted.

## IV. ADAPTIVE PLL FOR THE TRANSIENT STABILITY ENHANCEMENT

### A. General Idea

From the analysis in Section III-C, the LOS of the VSC with single equilibrium point during the fault is inevitable as long as $K_i > 0$. However, by setting $K_i$ equal to 0, (12) can be simplified as $\dot{\delta} = \dot{\delta}_1 = K_p\left(-I_{max}R_{line} - V_{gcpfault}\sin\delta\right)$, and consequently $\dot{\delta} = 0$ will always hold at the SEP. The transient stability of the VSC can thus be guaranteed as long as the equilibrium point exists.

However, with $K_i = 0$, the SRF-PLL becomes a first-order PLL [30], as shown in Fig. 11. The first-order PLL does mitigate the LOS problem during grid faults as long as equilibrium points exist, yet it suffers from the steady-state phase-tracking error when the grid frequency deviates from its nominal value [23]. To tackle this challenge, an adaptive PLL is introduced for enhancing the transient stability of the VSC. The basic idea of this method is to make the VSC operating with the SRF-PLL during the steady-state operation to achieve the zero phase-tracking error, and the SRF-PLL is switched to the first-order PLL only during grid fault-occurring/clearing process, which thus guarantees the transient stability of the VSC.

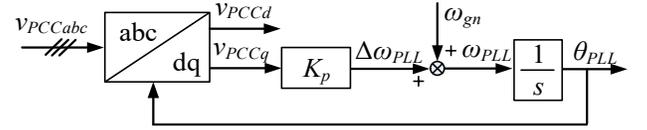

Fig. 11. Block diagram of the first-order PLL.

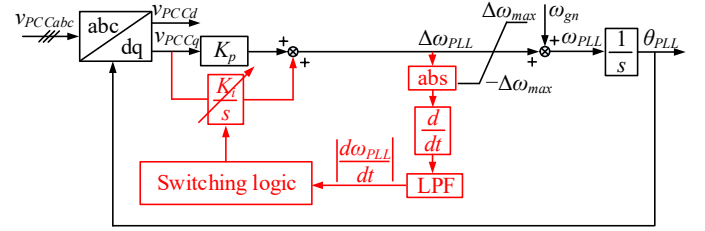

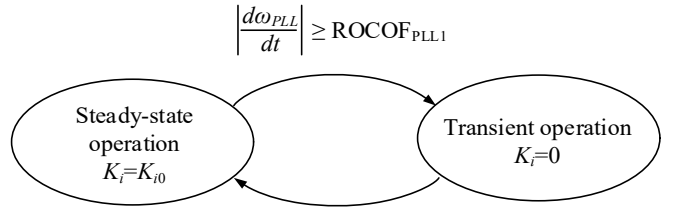

(b)

Fig. 12. Adaptive-PLL for the transient stability enhancement. (a) Control block diagram. (b) Mode switching logic.

Fig. 12(a) illustrates the control diagram of the adaptive PLL, and its switching logic is given in Fig. 12(b). Since $v_{PCCq}$ has an abrupt change during large grid disturbances, leading to an abrupt change of $\Delta\omega_{PLL}$ detected by the PLL, the integral gain $K_i$ can thus be changed based on the rate of change of frequency (ROCOF) detected by the PLL, i.e.,

$$K_i = 0, \qquad \left|\frac{d\omega_{PLL}}{dt}\right| \geq \text{ROCOF}_{PLL1}$$
$$K_i = K_{i0}, \qquad \left|\frac{d\omega_{PLL}}{dt}\right| < \text{ROCOF}_{PLL2} \quad (15)$$

where $K_{i0}$ is the designed integral gain of the PLL during the steady state operation, $\Delta\omega_{max}$ denotes the output frequency limit of the PLL. ROCOF$_{PLL1}$ and ROCOF$_{PLL2}$ represent the threshold values of ROCOF for switching the PLL between two different modes, respectively. $|d\omega_{PLL}/dt|$ is the ROCOF detected by the PLL, which is obtained by applying derivation to the absolute value of its output frequency, and a first-order low-pass filter (LPF) is added after the derivation in order to attenuate the high-frequency noise, i.e.,

$$G_{LPF}(s) = \frac{1}{T_{filter}s + 1}. \quad (16)$$

where $T_{filter}$ is the time constant of the LPF. $T_{filter}$ =180ms~240ms is recommended in [31], and $T_{filter}$=200ms is selected in this paper.

It should be noted that the frequency detected by the PLL can be deviated from the real grid frequency at the fault instant

[32]. Hence, the adaptive PLL proposed in this work does not rely on the derivative of the real grid frequency for the fault detection. In contrast, the high $|d\omega_{PLL}/dt|$ detected by the PLL essentially represents an abrupt change of $v_{PCCq}$, which indicates the large grid disturbances and dictates the mode switching, as shown in Fig. 12.

### B. Selection of ROCOF$_{PLL1}$ and ROCOF$_{PLL2}$

In order to adaptively switch the SRF-PLL to the first-order PLL during grid transients, it is known from (15) that ROCOF$_{PLL1}$ should be selected smaller than $|d\omega_{PLL}/dt|$ detected by the PLL at the fault instant. Therefore, the upper boundary of ROCOF$_{PLL1}$ can be given by

$$\text{ROCOF}_{PLL1} \leq \left|\frac{d\omega_{PLL}}{dt}\right|_{fault}$$
$$\approx \left(1-e^{-\Delta t/T_{filter}}\right)\left|\left(\frac{\dot{\delta}_{fault}-\dot{\delta}_{prefault}}{\Delta t}\right)\right| = \left(1-e^{-\Delta t/T_{filter}}\right)\left|\frac{\dot{\delta}_{fault}}{\Delta t}\right|. \quad (17)$$

where $\dot{\delta}_{prefault}$ is the frequency deviation of the PLL before the fault, which is zero. $\dot{\delta}_{fault}$ is the frequency deviation of the PLL at the fault instant, which can be calculated based on (12). $\Delta t$ represents the time duration of the fault transient, which is usually less than half cycle of the grid voltage, i.e., $\Delta t \leq 10$ ms [33]. $1-e^{-\Delta t/T_{filter}}$ represents the dynamic response of the LPF used in Fig. 12(a) [34].

Substituting (12) into (17), which yields

$$\text{ROCOF}_{PLL1} \leq \left|\frac{d\omega_{PLL}}{dt}\right|_{fault}$$
$$\approx \left(1-e^{-\Delta t/T_{filter}}\right)\frac{(K_p+K_i\int)\left|(-I_{max}R_{line}-V_{gcpfault}\sin\delta)\right|}{\Delta t} \quad (18)$$
$$= \left(1-e^{-\Delta t/T_{filter}}\right)\frac{(K_p+K_i\int)(I_{max}R_{line}+V_{gcpfault}\sin\delta)}{\Delta t}$$

Since $\delta>0$ always holds at the fault instant, the sufficient condition of (18) is derived as

$$\text{ROCOF}_{PLL1} \leq \left(1-e^{-\Delta t/T_{filter}}\right)\frac{K_p I_{max}R_{line}}{\Delta t}. \quad (19)$$

Substituting (10) into (19), which yields

$$\text{ROCOF}_{PLL1} \leq \frac{9.2 I_{max}R_{line}}{V_{gn}t_s\Delta t}\left(1-e^{-\Delta t/T_{filter}}\right). \quad (20)$$

The typical value of $R_{line}$ is 0.02 pu for the transmission grid [27], while this value is much larger in the distribution grid [27]. Nevertheless, the minimum value in the right-hand side of (20) is of concern to determine the upper boundary of ROCOF$_{PLL1}$, and thus $R_{line}$ = 0.02 pu is selected in this work. The typical settling time of the PLL is 100ms, i.e., $t_s$=100ms [24]. Substituting $V_{gn}$=1pu, $I_{max}$=1pu, $R_{line}$ = 0.02 pu, $t_s$=100ms, $\Delta t$ =10 ms into (20), which yields ROCOF$_{PLL1} \leq$ 8.8 Hz/s.

The lower boundary of ROCOF$_{PLL1}$ is determined based on the criterion of avoiding the adaptive PLL to be wrongly switched to the first-order PLL during the steady state operation. It is noted that the frequency fluctuation always exists in the real power grid [27]. The power system with a high penetration of renewable energy resources usually experiences a higher ROCOF [27], and the 2.5 Hz/s ROCOF withstand capability of the generation unit is specified in [35], which requires ROCOF$_{PLL1} \geq$ 2.5 Hz/s. Therefore, ROCOF$_{PLL1}$ is selected in the range between 2.5Hz/s and 8.8 Hz/s. In this paper, ROCOF$_{PLL1}$ = 5 Hz/s is selected in simulations and experimental tests.

When the adaptive PLL is switched to the first-order PLL at the fault instant, the system can always be stabilized at the new equilibrium point, after which the first-order PLL can be switched back to the SRF-PLL. The value of $|d\omega_{PLL}/dt|$ converges to zero at the equilibrium point in theory, but noises always exist in practice. Therefore, a small positive value of ROCOF$_{PLL2}$ is selected to enhance the robustness of the algorithm. In this work, ROCOF$_{PLL2}$ = 0.5 Hz/s is adopted in simulations and experimental tests.

It should be noted that if there is a grid frequency deviation during the fault, the steady-state phase tracking error will be inevitable when the adaptive-PLL switches to the first-order PLL. Consequently, the accuracy of the reactive current injection is also affected. However, this phase tracking error can be compensated after the adaptive-PLL switches back to the SRF-PLL (second-order PLL). The smaller time constant of the LPF can improve the dynamic of the switching between the first-order PLL and the SRF-PLL, but it also jeopardizes the robustness of the switching logic against noises. As will be shown in the simulation and experimental results, by using the LPF with the time constant of 200ms, the adaptive PLL is kept as the SRF-PLL in most of the fault period, where the accurate reactive current injection can be guaranteed.

### C. Comparative Analysis of Transient Stability of VSCs with Different PLLs

In this part, the transient stability of the VSC with different PLLs are compared with phase portraits. The main circuit parameters of the system are given in Table I. The lowest voltage that theoretically guarantees the transient stability of the system can be calculated as $V_{gcpfaultmin}=I_{max}R_{line}$=0.1 pu, i.e., when the voltage drops below 0.1 p.u., the PLL does not have equilibrium points and the LOS is inevitable.

TABLE I
MAIN CIRCUIT PARAMETERS USED IN SIMULATIONS

| SYMBOL | DESCRIPTION | VALUE (P.U.) |
|---|---|---|
| $V_{gcprms}$ | RMS value of the GCP voltage | 33 kV (1 p.u.) |
| $P$ | Power rating of the VSC | 1 MW (1 p.u.) |
| $f_g$ | Grid frequency | 50 Hz (1 p.u.) |
| $L_f$ | Inductance of the output filter | 0.096 p.u. |
| $L_{line}$ | Line inductance | 0.28 p.u. |
| $R_{line}$ | Line resistance | 0.1 p.u. |

TABLE II
CONTROL PARAMETERS OF THE PLL

| SYMBOL | CASE I | CASE II | CASE III |
|---|---|---|---|
| PLL structure | SRF-PLL | SRF-PLL | Adaptive-PLL |
| $\zeta$ | 0.5 | 1.5 | 1.5 |
| $t_s$ | 0.1s | 0.1s | 0.1s |

Three different cases with different PLL parameters, which are summarized in Table II, are compared. In cases I and II, the SRF-PLL is adopted during the normal operation and grid faults, and the difference between them is that the damping ratio is set as 0.5 in the case I and 1.5 in the case II. In the case III, besides the parameters used in the case II, the adaptive PLL shown in Fig. 12 is further employed to enhance the transient stability of the system.

The large-signal nonlinear responses of different PLLs are analyzed by using phase portraits, which are plotted based on the dynamic equation given in (11). Different depths of voltage sags are evaluated and two typical operating scenarios are considered, as shown in Fig. 13, where the points $a$ and $c$ represent the SEPs of the PLL before and after the fault, respectively. The system is stable if its phase portrait is converged to the new SEP $c$ after the fault, and it becomes unstable if its phase portrait is diverged.

Fig. 13 (a) shows the phase portraits of different PLLs when the GCP voltage drops to 0.14 pu. In this operating scenario, it can be calculated that $|v_{zq}|/|V_{gcpfault}|=0.1/0.14=0.71$, and it is known from Fig. 9 that the corresponding critical damping ratio is 0.695. Consequently, the LOS is inevitable in the case I, due to $\zeta=0.5<0.695$, as shown by the dashed dotted line in Fig. 13(a). In contrast, the synchronization can still be remained in the case II, where $\zeta=1.5>0.695$, as shown by the dashed line in Fig. 13(a). Hence, it is clear that the transient stability of the system can be enhanced by the increased damping ratio of the PLL in this operating scenario. However, when the GCP voltage drops to the theoretically lowest voltage limit, i.e., 0.10 p.u., the system has only one equilibrium point and the system becomes unstable with the PLL parameters in both cases I and II. In contrast, the adaptive PLL can stabilize the system in both operating scenarios, as indicated by the solid lines in both Fig. 13(a) and Fig. 13(b).

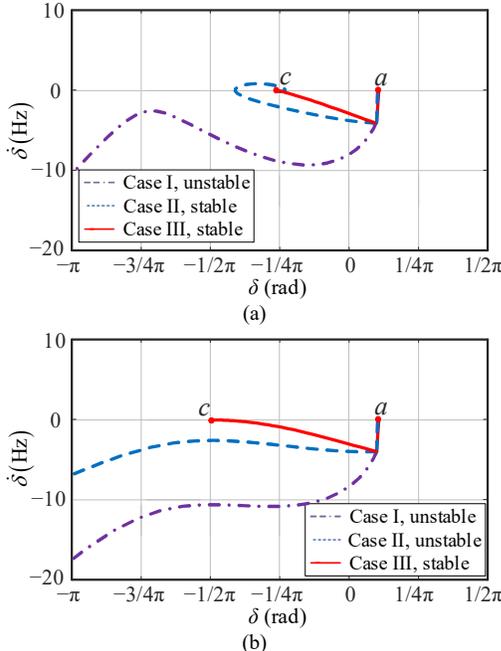

Fig. 13. Phase portraits of the PLL with different design when the GCP voltage drops. (a) $V_{gcp}$ drops to 0.14 p.u. (b) $V_{gcp}$ drops to 0.10 p.u.

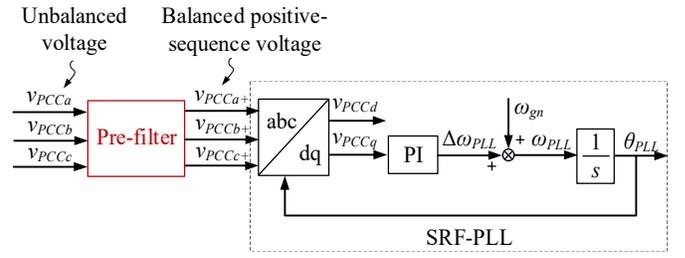

Fig. 14. Block diagram of the SRF-PLL with a pre-filter used to extract the positive-sequence voltage.

### D. Asymmetrical Faults

It is noted that only the symmetrical faults have been considered in the above analysis. Yet, the asymmetrical faults, which introduce both positive- and negative-sequence voltages to the VSC, are more commonly seen in practice, and the zero-sequence voltage does not appear due to the use of the delta-wye transformers for VSCs [27].

During asymmetrical faults, a pre-filter is generally used at the input of SRF-PLL for extracting the positive-sequence voltage, as shown in Fig. 14 [36]-[38]. However, the pre-filter is designed to have little effect on the synchronization stability of the SRF-PLL. This is because the pre-filter needs to detect the positive-sequence voltage with a much faster speed than the synchronization dynamic of the SRF-PLL during the fault instant [24], such that there are no interactions between the sequence component detection and the grid synchronization [36].

Hence, the assumption that the input voltage of the SRF-PLL is three-phase balanced still holds during asymmetrical faults, thanks to the fast positive-sequence voltage detection of the pre-filter. The performed analysis in this paper is thus valid, and same conclusions, i.e., the transient stability of the VSC can be improved by increasing the damping ratio of the SRF-PLL, and it can be further enhanced by using the first-order PLL, can also be drawn for asymmetrical faults. This statement will be further justified by the simulation and experiments performed in the next section.

## V. SIMULATION AND EXPERIMENTAL RESULTS

### A. Simulation Results

To validate the theoretical analysis, time-domain simulations are carried out in the MATLAB/Simulink and PLECS blockset with the nonlinear switching circuit model shown in Fig. 1. The parameters given in Table I and Table II are adopted. The output frequency limits of the PLL are set as 45 Hz~55 Hz. In the normal operation, the VSC operates with $I_d=I_{max}$, $I_q=0$. During severe grid faults, the VSC injects the rated reactive current into the grid to support the grid voltage, i.e., $I_d=0$, $I_q=-I_{max}$.

Moreover, to further highlight the advantage of the proposed method, the method that freezes the PLL during the fault [16] is also simulated. In this method, the phase angle used with the VSC after freezing ($\theta_{freeze}$) is determined by the measured phase angle and frequency of the PLL before freezing ($\theta_{prefreeze}$ and $\omega_{prefreeze}$), i.e., $\theta_{freeze} = \theta_{prefreeze} + \omega_{prefreeze} \cdot t$.

Corresponding to the scenario considered in Fig. 13(a), Fig. 15 shows the simulation results of the VSC during the symmetrical fault, where three-phase voltages drop to 0.14 p.u at $t$=2.5s, and the fault is cleared at $t$=3.1s. It is clear that the PLL with the parameter $\zeta$=0.5 (case I) cannot remain synchronization with the grid during the fault, and the output frequency of the PLL is saturated at its lower limit (45Hz), which cannot be recovered to the grid frequency, leading to a diverged $\delta$, as shown in Fig. 15 (a). However, the PLL with the parameter $\zeta$=1.5 (case II) and the adaptive PLL (case III) can still be kept synchronized with the grid during the fault, as shown in Fig. 15 (b) and (c). The simulation results agree well with the phase portrait analyses in Fig. 13(a).

It is worth mentioning that $K_i$ of the adaptive-PLL only switches to zero during the grid transients, i.e. the fault-occurring/-clearing instants, rather than the whole fault period. As shown in Fig. 15 (c), $K_i$ switches back to its designed value when the VSC reaches to a new steady state during the fault, which implies that the first-order PLL is switched back to the SRF-PLL. Thus, the accurate phase tracking of the PLL can be guaranteed even during the fault period. Moreover, the seamless transfer between the normal operation and the transient operation is also achieved with the adaptive PLL.

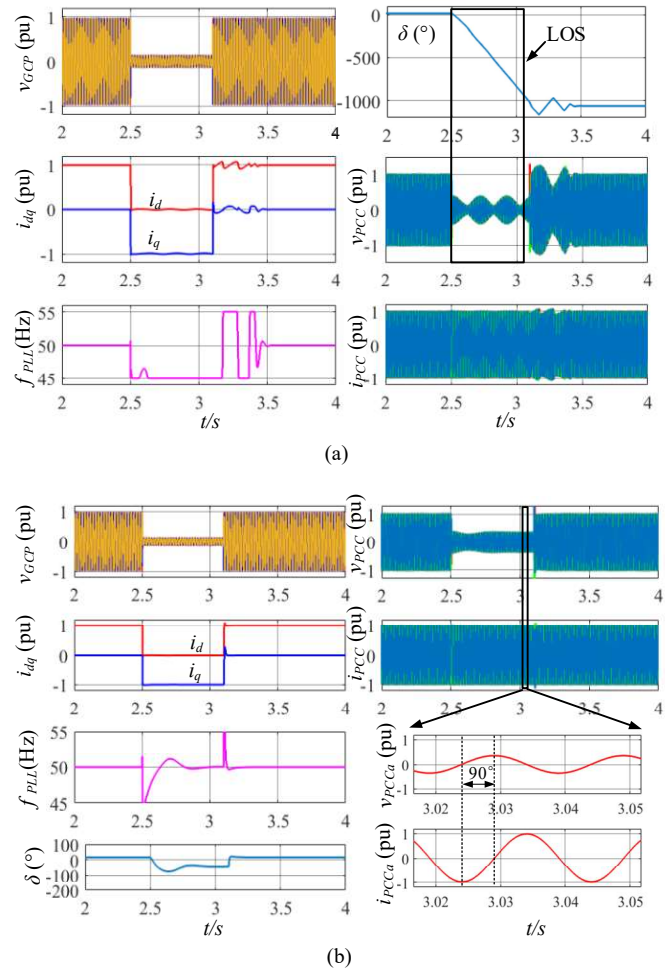

(a)

(b)

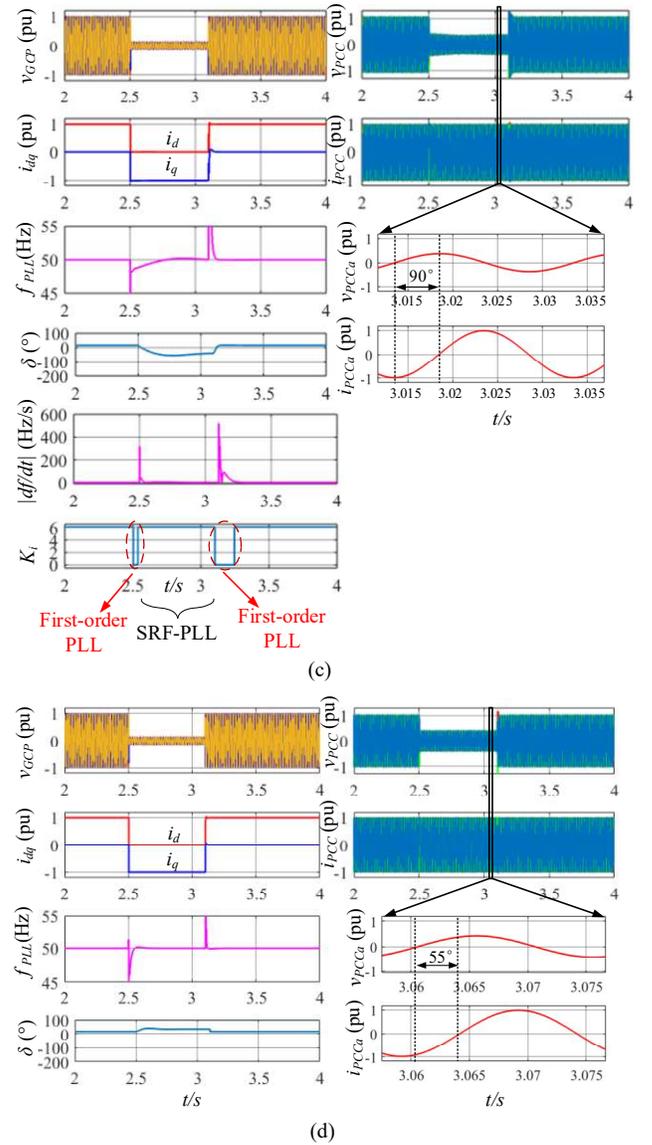

(c)

(d)

Fig. 15. Simulation results of the VSC during the symmetrical fault, where $V_{gcp}$ drops to 0.14 p.u. (a) Case I: PLL with $\zeta$=0.5, unstable. (b) Case II: PLL with $\zeta$=1.5, stable and accurate phase angle detection. (c) Case III: Adaptive PLL, stable and accurate phase angle detection. (d) Freezing the PLL: stable but inaccurate phase angle detection

On the other hand, if the PLL is activated during the fault, the accurate phase angle detection can always be guaranteed if there is no LOS. As shown in Fig. 15 (b) and (c), the phase difference between the PCC voltage and the injected current during the fault is 90º, indicating the pure reactive current injection. In contrast, by freezing the PLL, the phase angle is not affected by the grid voltage dip and thus the stability can be remained, but it fails to detect the correct grid phase angle. As shown in Fig. 15 (d), the phase difference between the PCC voltage and the injected current during the fault is 55º, indicating the VSC fails to inject the right amount of reactive current, which violates the grid code.

Fig. 16 shows the simulation results of the VSC during the symmetrical fault, where three-phase voltages drop to 0.10 p.u at $t$=2.5s, which corresponds to the operating scenario in Fig. 13(b), and the fault is cleared at $t$=3.1s. Since there is only one equilibrium point during the fault, only the VSC with the adaptive PLL can be kept synchronized in this scenario, as shown in Fig. 16 (a)-(c). The simulation results confirm the phase portrait analyses provided in Fig. 13 (b). Moreover, by using the adaptive PLL, not only the transient stability, but also the accurate phase angle detection, can be guaranteed. As shown in Fig. 16 (c), the phase difference between the PCC voltage and the injected current during the fault is 90º, indicating a purely reactive current injection. Similarly, by freezing the PLL, the stability can be guaranteed but the VSC fails to detect the right grid phase angle, the phase difference between the PCC voltage and the injected current during the fault is 64º, as shown in Fig. 16 (d). This indicates that the VSC fails to inject the rated reactive current if the PLL is frozen.

Figs. 17-18 further show the simulation results of the VSC during asymmetrical faults. The dual second-order generalized integrator (SOGI) based pre-filter is adopted [38], and the parameters of the SRF-PLL given in Table II are used. The magnitudes of the positive-sequence voltages during asymmetrical faults are selected to be same as that used in the phase portrait analysis given by Fig. 13, which are 0.14 pu and 0.10 pu, respectively.

Corresponding to the scenario considered in Fig. 13(a), Fig. 17 shows the simulation results of the VSC under the asymmetrical fault with $V_{gcpa}$=0.42 pu, and $V_{gcpb}$=$V_{gcpc}$= 0 pu, which corresponds to 0.14 pu positive-sequence voltage. It is clear that LOS occurs when the PLL with the parameter $\zeta$=0.5 (case I) is adopted, as shown in Fig. 17(a). In contrast, the system can be kept stable by increasing the damping ratio of the PLL ($\zeta$=1.5, case II) or using the adaptive PLL (case III), as shown in Figs. 17 (b) and (c). The simulation results agree well with the phase portrait analysis shown in Fig. 13 (a).

Corresponding to the scenario considered in Fig. 13(b), Fig. 18 shows the simulation result of the VSC under the asymmetrical fault with $V_{gcpa}$=0.3 pu, and $V_{gcpb}$=$V_{gcpc}$= 0 pu, which corresponds to 0.10 pu positive-sequence voltage. It is clear that the system can be kept stable only with the adaptive PLL, as shown in Fig. 18 (c), which also agrees with the phase portrait analysis shown in Fig. 13 (b).

Hence, the simulation results in Figs. 17-18 justify that the performed analysis also applies to asymmetrical faults when the pre-filtered PLL is used, and the same findings can also be drawn, i.e. the transient stability of the VSC can be improved by increasing the damping ratio of the SRF-PLL, and it can be further enhanced by using the first-order PLL.

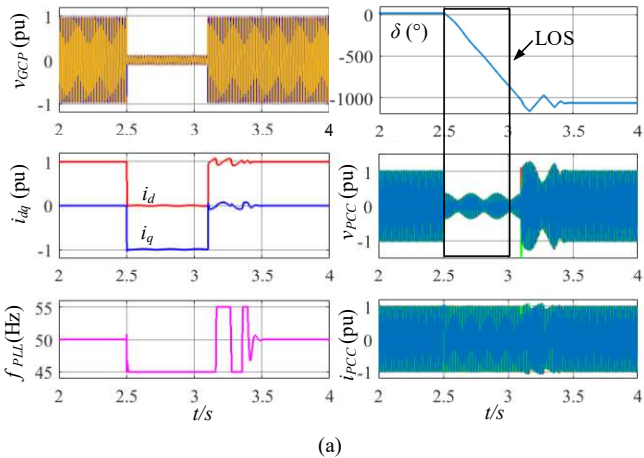

(a)

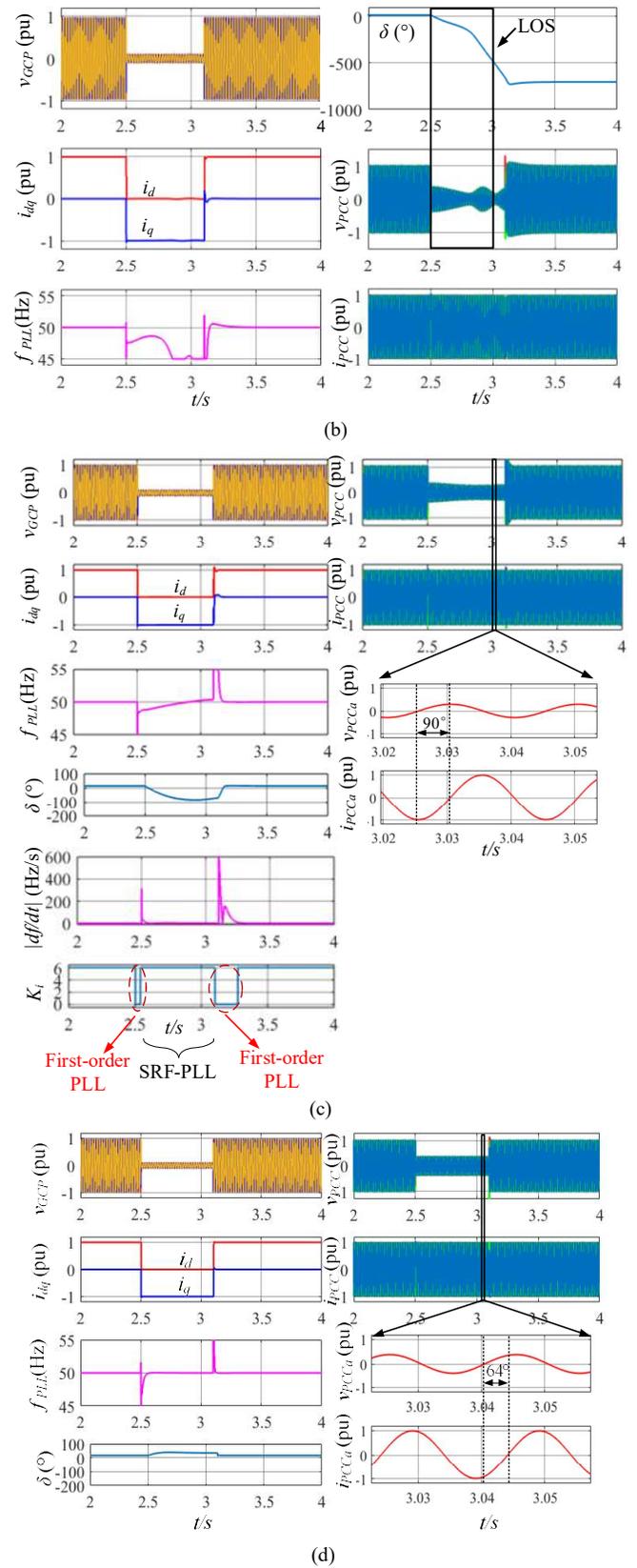

Fig. 16. Simulation results of the VSC during the symmetrical fault, where $V_{gcp}$ drops to 0.10 p.u. (a) Case I: PLL with $\zeta$=0.5, unstable. (b) Case II: PLL with $\zeta$=1.5, unstable. (c) Case III: Adaptive PLL, stable and accurate phase angle detection. (d) Freezing PLL: stable but inaccurate phase angle detection.

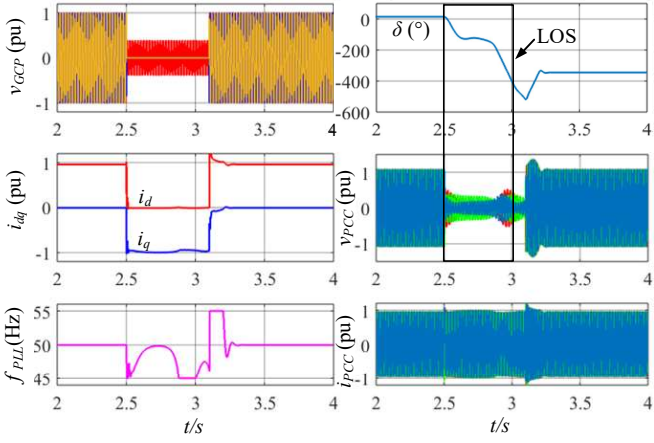
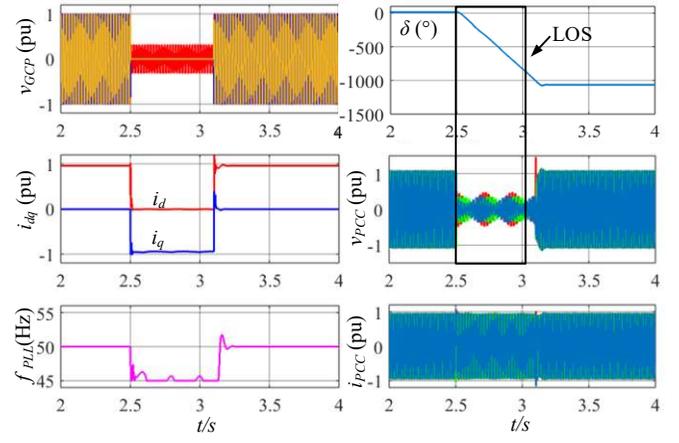
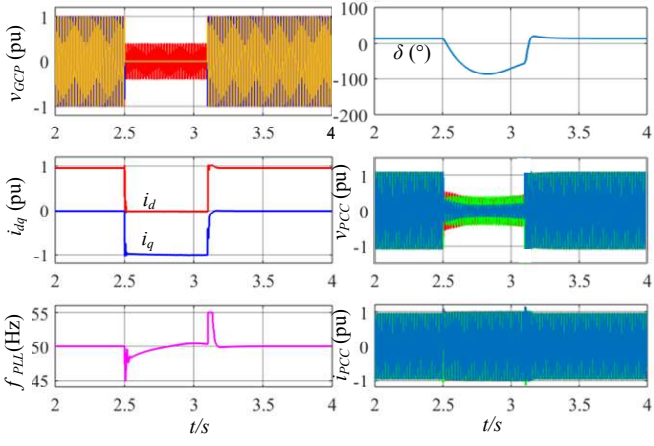
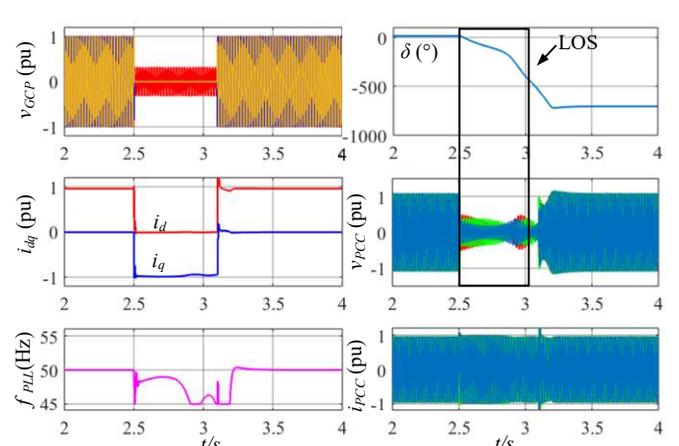
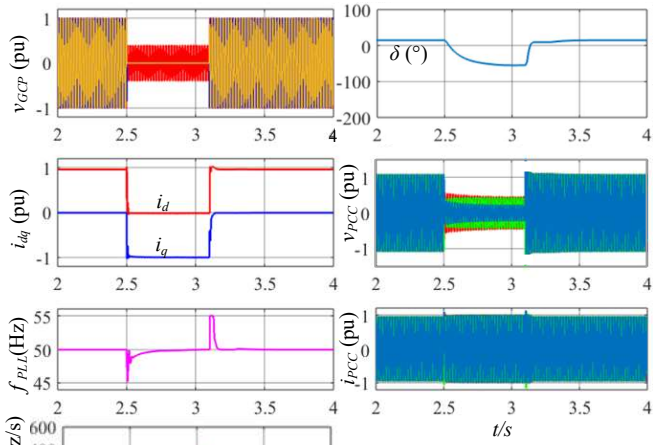
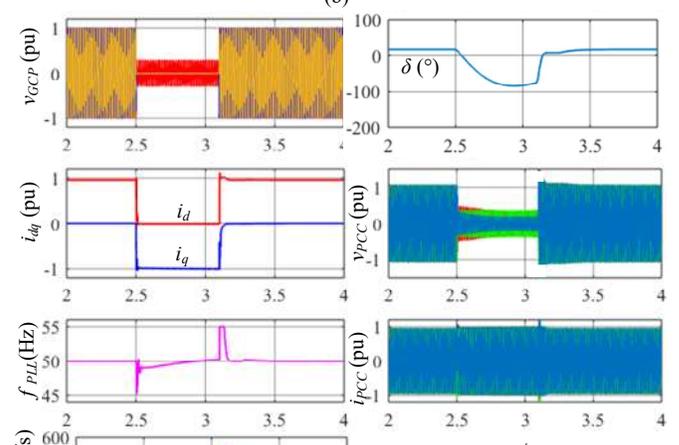
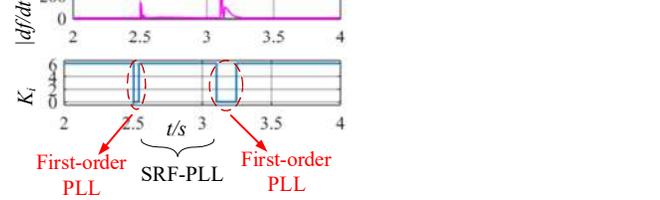
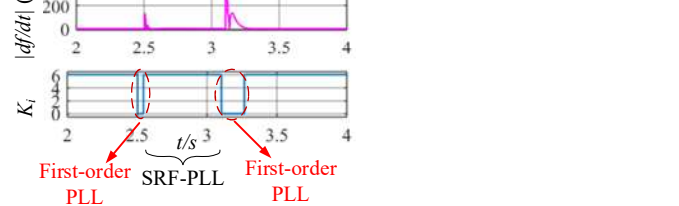

Fig. 17. Simulation results of the VSC during the asymmetrical fault, where $V_{gcpa} = 0.42$ p.u. and $V_{gcpb} = V_{gcpc} = 0$ p.u. (a) Case I: PLL with the parameter $\zeta=0.5$, unstable. (b) Case II: PLL with the parameter $\zeta=1.5$, stable. (c) Case III: Adaptive PLL, stable.

Fig. 18. Simulation results of the VSC during the asymmetrical fault, where $V_{gcpa} = 0.3$ p.u. and $V_{gcpb} = V_{gcpc} = 0$ p.u. (a) Case I: PLL with the parameter $\zeta=0.5$, unstable. (b) Case II: PLL with the parameter $\zeta=1.5$, unstable. (c) Case III: Adaptive PLL, stable.

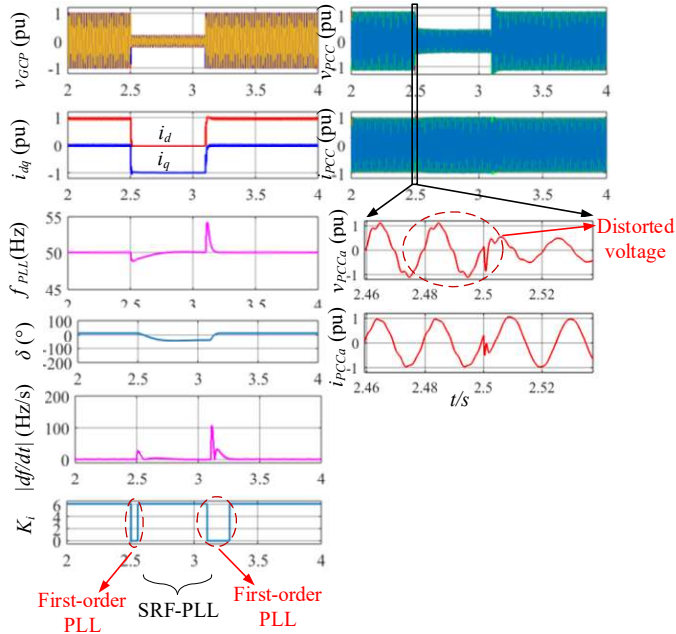

Fig. 19. Simulation results of the VSC with the adaptive PLL during the symmetrical fault. $V_{gcp}$ drops to 0.2 pu and the grid voltage includes 5% of 5$^{th}$-order harmonic and 8% of 7$^{th}$-order harmonic.

Fig. 19 shows the simulation result of the VSC with the adaptive PLL under the distorted grid voltage (including 5% of 5$^{th}$- order harmonic and 8% of 7$^{th}$- order harmonic). It is clear that neither the switching logic nor the grid phase angle detection accuracy is affected by the grid harmonics, indicating the strong robustness of the proposed adaptive-PLL against grid harmonics.

### B. Experimental Results

To further verify the simulation results, the experimental tests are carried out with a three-phase grid-connected converter with the downscaled voltage and power ratings. However, the per unit values of parameters used in the experiment are the same as that used in the simulation, which are listed in Table III. The experimental setup is shown in Fig. 20. The circuit tested in the experiment is identical to that shown in Fig. 1. The control algorithm is implemented in the DS1007 dSPACE system, where the DS5101 digital waveform output board is used for generating the switching pulses, and the DS2004 high-speed A/D board is used for the voltage and current measurements. The active/reactive current, the output frequency and the integral gain of the PLL are outputted through the DS2102 high-speed D/A board. A constant dc voltage supply is used at the dc-side, and a 45 kVA Chroma 61850 grid simulator is used to generate the grid voltage.

Figs. 21 shows the measured results of the VSC during the symmetrical fault, where three-phase voltages drop to 0.14 pu. Three different PLLs listed in Table II are compared. The VSC operates with $I_d=I_{max}$, $I_q=0$ during the normal operation and $I_d=0$, $I_q=-I_{max}$ when $V_{gcp}$ drops more than 50% of the nominal voltage. It can be seen from Fig. 21 (a) that the LOS of the system takes place when the PLL with parameter $\zeta=0.5$ is used, where the output frequency of the PLL is saturated at the lower limit (45Hz) and cannot be recovered to the nominal grid frequency during the fault. Moreover, a large frequency swing can be observed before the system is re-synchronized with the grid during the grid voltage recovery. In contrast, both the PLL with $\zeta=1.5$ and the adaptive-PLL can be kept stable in this scenario, as shown in Fig. 21 (b) and (c). These test results confirm the theoretical analysis and the simulation results in Figs. 13 (a) and 15.

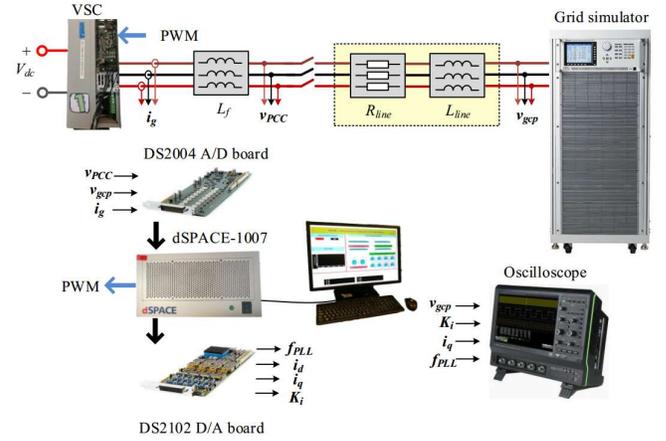

Fig. 20. Configuration of the experimental setup

TABLE III
MAIN CIRCUIT PARAMETERS USED IN EXPERIMENTS

| SYMBOL | DESCRIPTION | VALUE (P.U.) |
|---|---|---|
| $V_{gcprms}$ | RMS value of the GCP voltage | 110 V (1 p.u.) |
| $P$ | Power rating of the VSC | 3.6 kW (1 p.u.) |
| $f_g$ | Grid frequency | 50 Hz (1 p.u.) |
| $L_f$ | Inductance of the output filter | 0.096 p.u. |
| $L_{line}$ | Line inductance | 0.28 p.u. |
| $R_{line}$ | Line resistance | 0.1 p.u. |

Figs. 22 shows the measured results of the VSC during the symmetrical fault, where three-phase voltages drop to 0.10 pu. Three different PLLs listed in Table II are compared. It is clear that only the VSC with the adaptive PLL can be kept synchronized with power grid, thanks to the fact that $K_i$ of the adaptive PLL can be automatically switched to zero during the fault-occurring and fault-clearing transients, as shown in Fig. 22 (c). The experimental tests match well with the theoretical predictions and the simulation case studies in Figs. 13(b) and 16.

Corresponding to the simulation studies carried out in Figs 17 and 18, Figs. 23 and 24 show the measured results of the VSC under asymmetrical faults. For the asymmetrical fault with $V_{gcpa}=0.42$ pu, $V_{gcpb}=V_{gcpc}=0$ pu, the LOS occurs when the PLL with the parameter $\zeta=0.5$ (case I) is adopted, as shown in Fig. 23 (a). Yet, the system can be kept stable by increasing the damping ratio of the PLL ($\zeta=1.5$, case II) or using the adaptive PLL (case III), as shown in Figs. 23 (b) and (c). Moreover, when $V_{gcpa}$ is further dropped to 0.3 pu, only the adaptive PLL can stabilize the system, as shown in Fig. 24 (c). The experimental results match well with the simulation studies.

Fig. 25 shows the measured result of the VSC with the adaptive PLL under the distorted grid voltage (including 5% of $5^{th}$- order harmonic and 8% of $7^{th}$- order harmonic), and the satisfactory performance of the adaptive-PLL can be observed, which further confirms the simulation results shown in Fig. 19.

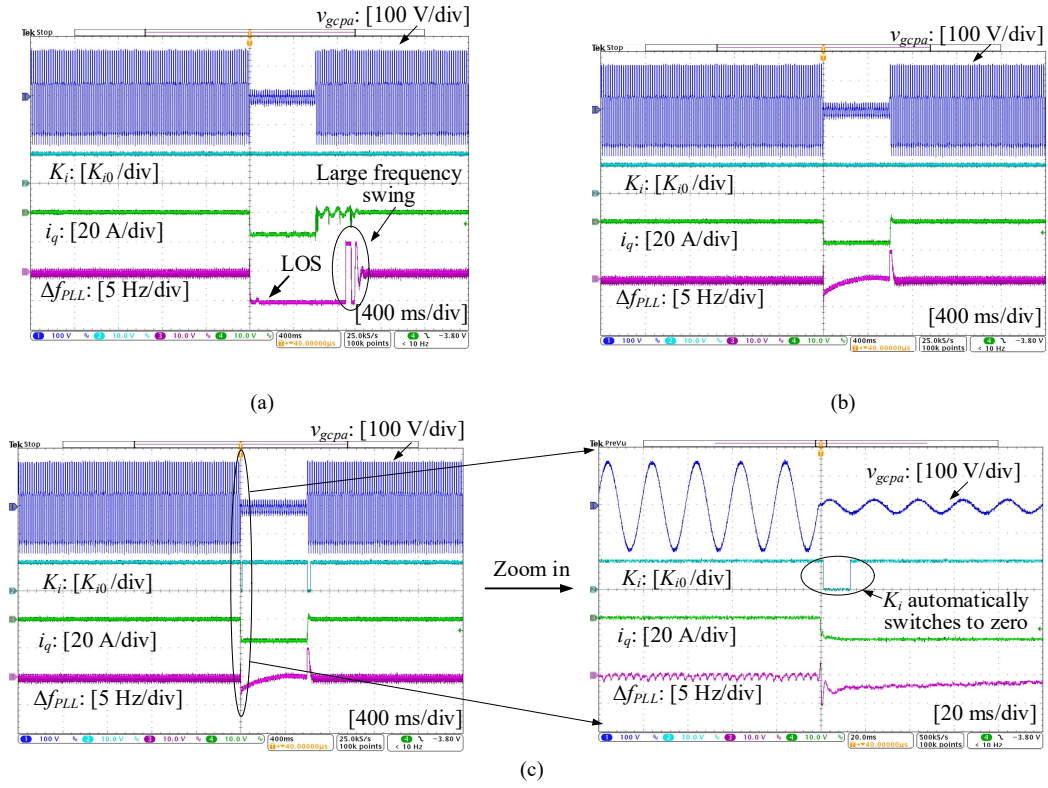

Fig. 21. Experimental results of the VSC with different designed PLLs during the symmetrical fault, where $V_{gcp}$ drops to 0.14 p.u. (a) Case I: PLL with $\zeta=0.5$, unstable. (b) Case II: PLL with $\zeta=1.5$, stable. (c) Case III: Adaptive PLL, stable.

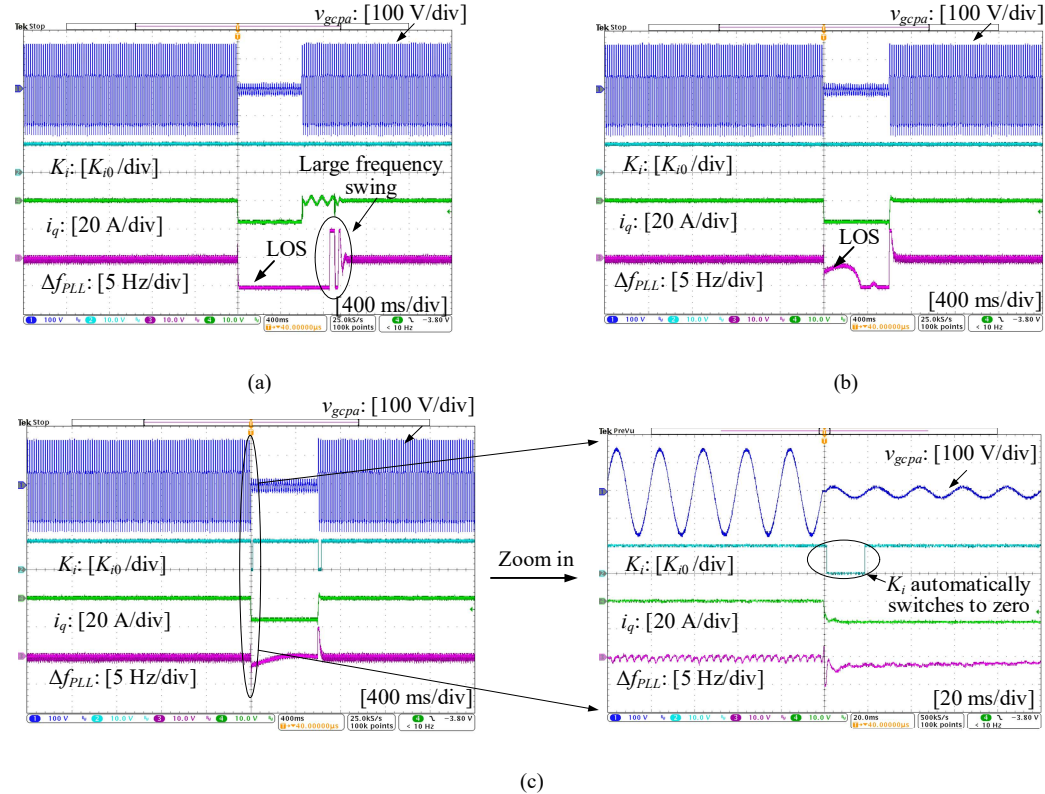

Fig. 22. Experimental results of the VSC with different designed PLLs during the symmetrical fault, where $V_{gcp}$ drops to 0.10 p.u. (a) Case I: PLL with $\zeta=0.5$, unstable. (b) Case II: PLL with $\zeta=1.5$, unstable. (c) Case III: Adaptive PLL, stable.

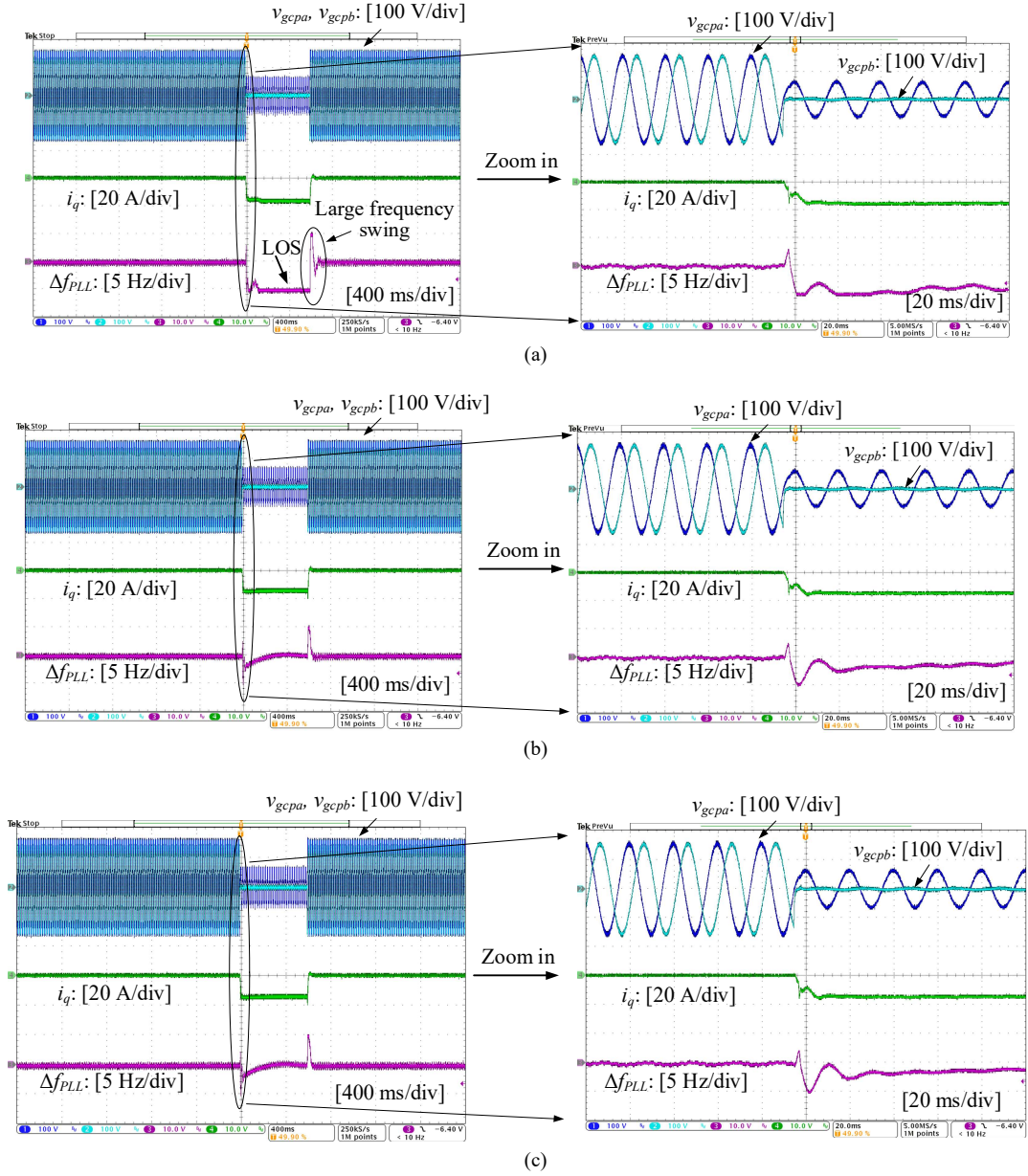

Fig. 23. Experimental results of the VSC during the asymmetrical fault, where $V_{gcpa} = 0.42$ p.u. and $V_{gcpb} = V_{gcpc} = 0$ p.u. (a) Case I: PLL with the parameter $\zeta=0.5$, unstable. (b) Case II: PLL with the parameter $\zeta=1.5$, stable. (c) Case III: Adaptive PLL, stable.

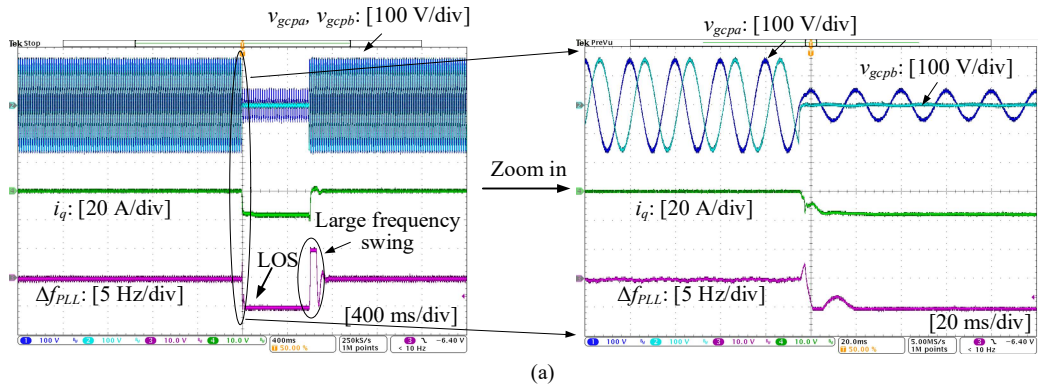

(a)

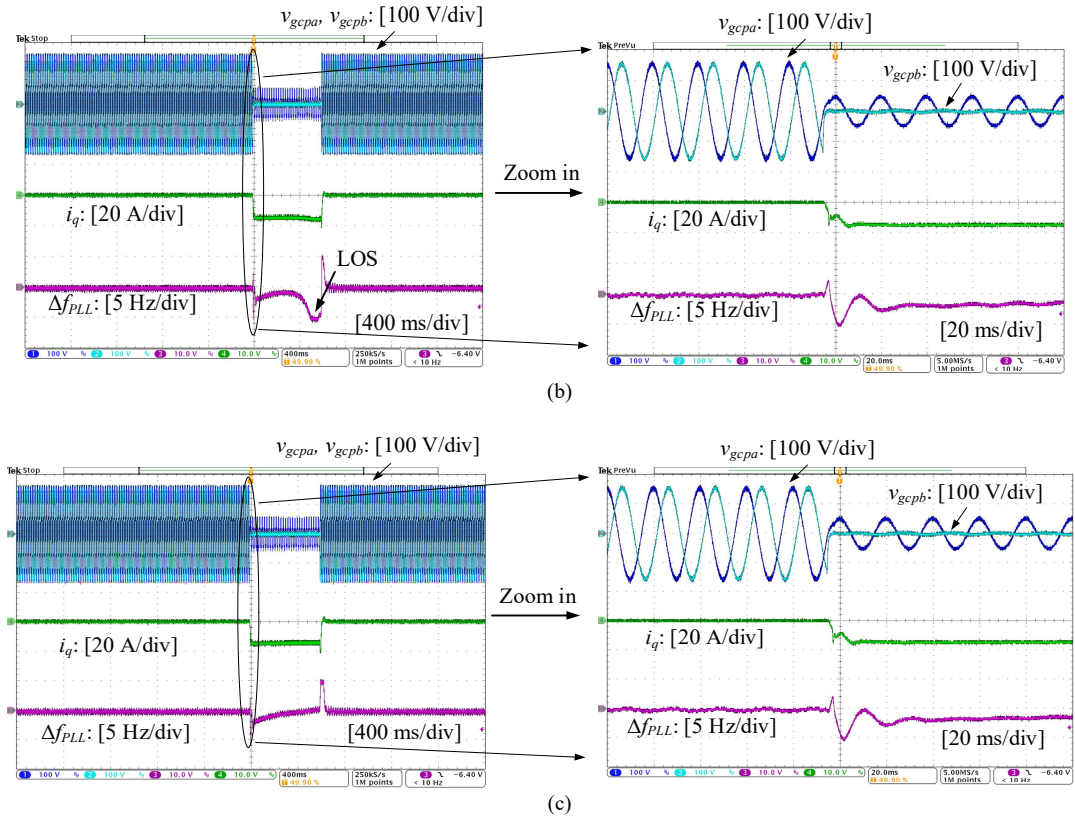

Fig. 24. Experimental results of the VSC during the asymmetrical fault, where $V_{gcpa} = 0.3$ p.u. and $V_{gcpb} = V_{gcpc} = 0$ p.u. (a) Case I: PLL with the parameter $\zeta=0.5$, unstable. (b) Case II: PLL with the parameter $\zeta=1.5$, unstable. (c) Case III: Adaptive PLL, stable.

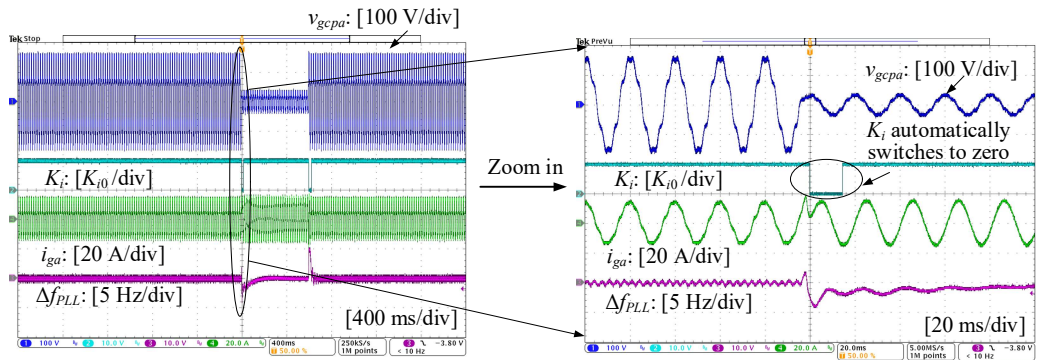

Fig. 25. Experimental results of the VSC with the adaptive PLL during the symmetrical fault. $V_{gcp}$ drops to 0.2 pu, and the grid voltage includes 5% of 5th –order harmonic and 8% of 7th –order harmonic.

## VI. CONCLUSION

This paper has analyzed the impact of the PLL on the transient stability of VSCs during grid faults. The large-signal nonlinear responses of the PLL with different parameters have been characterized by means of the phase portrait. The major findings of the paper are summarized as follows:

1) The transient stability of the VSC can be enhanced by increasing the damping ratio of the SRF-PLL when the system has two equilibrium points during the fault. The value of the critical damping ratio to stabilize the system is identified based on the voltage ratio $|v_{zq}|/|V_{gcpfault}|$. However, the LOS of the SRF-PLL is inevitable when only one equilibrium point exists during the fault.

2) In contrast to the SRF-PLL, the first-order PLL has no LOS problem whenever the system has equilibrium points, yet it suffers from the steady-state phase tracking error when the grid frequency has a steady-state drift from its nominal value.

3) The proposed adaptive PLL enables the VSC to operate with the SRF-PLL in the steady state operation and with the first-order PLL during the fault-occurring/-clearing transients, which not only guarantees the transient stability of the system, but also ensures the phase-tracking accuracy even during the grid fault.

All the findings have been elaborated theoretically and confirmed by time-domain simulations and experimental tests.

## VII. REFERENCES

[1] F. Blaabjerg, Y. Yang, D. Yang and X. Wang, "Distributed Power Generation Systems and Protection," in *Proceedings of the IEEE*, vol. 105, no. 7, pp. 1311-1331, July 2017.

[2] X. Wang and F. Blaabjerg, "Harmonic stability in power electronic based power systems: concept, modeling, and analysis," *IEEE Trans. Smart Grid.*, vol. 10, no. 3, pp. 2858–2870, May. 2019.

[3] L. Harnefors, M. Bongiorno, and S. Lundberg, "Input-admittance calculation and shaping for controlled voltage-source converters," *IEEE Trans. Ind. Electron.*, vol. 54, no. 6, pp. 3323-3334, Dec. 2007.

[4] B. Wen, D. Boroyevich, R. Burgos, P. Mattavelli, and Z. Shen, "Analysis of D-Q small-signal impedance of grid-tied inverters," *IEEE Trans. Power Electron.*, vol. 31, no. 1, pp. 675-687, Jan. 2016.

[5] X. Wang, L. Harnefors, and F. Blaabjerg, "Unified impedance model of grid-connected voltage-source converters," *IEEE Trans. Power Electron., IEEE Trans. Power Electron.*, vol. 33, no. 2, pp. 1775-1787, Feb. 2018.

[6] Joint NERC and WECC Staff Report, "900 MW fault induced solar photovoltaic resource interruption disturbance report," Atlanta, USA, Feb. 2018, [Online]. Available: www.nerc.com.

[7] L. Hadjidemetriou, E. Kyriakides, and F. Blaabjerg, "An adaptive tuning mechanism for phase-locked loop algorithms for faster time performance of interconnected renewable energy sources," *IEEE Trans. Ind. Appl.*, vol. 51, no. 2, pp. 1792–1804, March 2015.

[8] M. Karimi-Ghartemani, S. A. Khajehoddin, P. K. Jain, and A. Bakhshai, "Problems of startup and phase jumps in PLL systems," *IEEE Trans. Power Electron.*, vol. 27, no. 4, pp. 1830–1838, Apr. 2012.

[9] T. Thacker, D. Boroyevich, R. Burgos, and F. Wang, "Phase-locked loops noise reduction via phase detector implementation for single-phase systems," *IEEE Trans. Ind. Electron.*, vol. 58, no. 6, pp. 2482–2490, Jun. 2011.

[10] D. Dong, B. Wen, D. Boroyevich, P. Mattavelli, and Y. Xue, "Analysis of phase-locked loop low-frequency stability in three-phase grid-connected power converters considering impedance interactions," *IEEE Trans. Ind. Electron.*, vol. 62, no. 1, pp. 310–321, Jan. 2015.

[11] C. Zhang, X. Cai, and Z. Li, "Transient stability analysis of wind turbines with full-scale voltage source converter," *Proceedings of the CSEE.*, vol. 37, no. 14, pp. 4018–4026, July. 2017.

[12] H. Wu and X. Wang, "Transient stability impact of the phase-locked loop on grid-connected voltage source converters," *IEEE International Power Electronics Conference (IPEC-ECCE Asia)*, 2018.

[13] H. Wu and X. Wang, "An adaptive phase-locked loop for the transient stability enhancement of grid-connected voltage source converters," in *Proc. 2018 IEEE Energy Conversion Congress and Exposition (ECCE)*, 2018.

[14] M. Taul, X. Wang, P. Davari, and F. Blaabjerg, "An efficient reduced-order model for studying synchronization stability of grid-following converters during grid faults", in Proc. *IEEE COMPEL*, June 2019, in Press.

[15] Q. Hu, L. Fu, F. Ma, and F. Ji, "Large signal synchronizing instability of PLL-based VSC connected to weak ac grid," *IEEE Trans. Power Syst.*, vol. 34, no.4, pp. 3220–3229, July 2019.

[16] B. Weise, "Impact of k-factor and active current reduction during fault-ride-through of generating units connected via voltage-sourced converters on power system stability," *IET Renewable Power Generation*, vol. 9, no. 1, pp. 25–36, 2015.

[17] BDEW Technical Guideline, Generating Plants Connected to the Medium- Voltage Network [EB/OL], June 2008 issue.

[18] V. Diedrichs, A. Beekmann, and S. Adloff, "Loss of (angle) stability of wind power plants - the underestimated phenomenon in case of very low short circuit ratio," in *Wind Integration Workshop, 2011 Aarhus, Denmark*, October 2011.

[19] S. Ma, H. Geng, L. Liu, G. Yang, and B. C. Pal, "Grid-synchronization stability improvement of large scale wind farm during severe grid fault," *IEEE Trans. Power Syst.*, vol. 33, no.1, pp. 216–226, Jan 2018.

[20] O. Göksu, R. Teodorescu, C. L. Bak, F. Iov, and P. C. Kjær, "Instability of wind turbine converters during current injection to low voltage grid faults and PLL frequency based stability solution," *IEEE Trans. Power Syst.*, vol. 29, pp. 1683–1691, July 2014.

[21] H. Geng, L. Liu, and R. Li, "Synchronization and reactive current support of PMSG based wind farm during severe grid fault," *IEEE Trans. Sustain. Energy*, vol. 9, no. 4, pp. 1596–1604, Oct 2018.

[22] M. Taul, X. Wang, P. Davari and F. Blaabjerg, "An overview of assessment methods for synchronization stability of grid-connected converters under severe symmetrical grid faults," *IEEE Trans. Power Electron.*, early access, pp. 1–1, 2019.

[23] S. K. Chung, "A phase tracking system for three phase utility interface inverters," *IEEE Trans. Power Electron.*, vol. 15, no. 3, pp. 431-438, May. 2000.

[24] R. Teodorescu, M. Liserre, and P. Rodriguez, *Grid Converters for Photovoltaic and Wind Power Systems*, Wiley Press, 2011.

[25] Y. Gu, N. Bottrell, and T. C. Green, "Reduced-order models for representing converters in power system studies," *IEEE Trans. Power Electron.*, vol. 33, no. 4, pp. 3644–3654, 2018.

[26] L. Harnefors, X. Wang, A. Yepes, and F. Blaabjerg, "Passivity-based stability assessment of grid-connected VSCs - an overview," *IEEE Jour. Emer. Select. Top. Power Electron.*, vol. 4, no. 1, pp. 116-125, Mar. 2016.

[27] P. Kundur, *Power System Stability and Control*. New York, NY, USA: McGraw-Hill, 1994.

[28] H. Wu and X. Wang, "Design-oriented transient stability analysis of grid-connected converters with power synchronization control" *IEEE Trans. Ind. Electron.*, vol. 66, no. 8, pp. 6473–6482, Aug. 2019.

[29] Steven H. Strogatz, *Nonlinear Dynamics and Chaos: With Applications to Physics, Biology, Chemistry, and Engineering*. Perseus Books, 1994.

[30] F. M. Gardner, *Phaselock Techniques*, 3rd ed. Hoboken, NJ, USA: Wiley, 2005.

[31] European network of transmission system operators for electricity (ENTSO-E), "Frequency measurement requirements and usage", 2018.

[32] 1200 MW Fault Induced Solar Photovoltaic Resource Interruption Disturbance Report, NERC Report, June 2017, available online: www.nerc.com

[33] M. Baran, I El-Markaby, "Fault analysis on distribution feeders with distributed generators", *IEEE Trans. Power Syst.*, Vol. 2, No. 4, Nov. 2005.

[34] G. C. Goodwin, S. F. Graebe, and M. E. Salgado, *Control System Design*. Upper Saddle River, NJ, USA: Prentice-Hall, 2000.

[35] European network of transmission system operators for electricity (ENTSO-E), "Rate of change of frequency (RoCoF) withstand capability", 2018.

[36] X. Guo, W. Wu, and Z. Chen, "Multiple-complex coefficient-filter-based phase-locked loop and synchronization technique for three-phase grid-interfaced converters in distributed utility networks," *IEEE Trans. Ind. Electron.,* vol. 58, no. 4, pp. 1194–1204, Apr. 2011

[37] Y. Wang and Y. Li, "Grid synchronization PLL based on cascaded delayed signal cancellation," *IEEE Trans. Power Electron.*, vol. 26, no. 7, pp. 1987–1997, Jul. 2011.

[38] P. Rodríguez, A. Luna, R. S. Muñoz-Aguilar, I. Etxeberria-Otadui, R. Teodorescu, and F. Blaabjerg, "A stationary reference frame grid synchronization system for three-phase grid-connected power converters under adverse grid conditions," *IEEE Trans. Power Electron.*, vol. 27, no. 1, pp. 99–112, Jan. 2012.

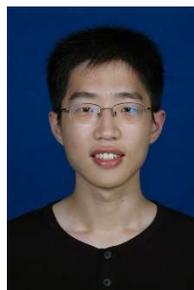

**Heng Wu** (S'17) received the B.S. degree in electrical engineering and the M.S. degree in power electronic engineering both from Nanjing University of Aeronautics and Astronautics (NUAA), Nanjing, China, in 2012 and 2015, respectively. He is currently working toward the Ph.D. degree in power electronic engineering in Aalborg University, Aalborg, Denmark.

From 2015 to 2017, He was an Electrical Engineer with NR Electric Co., Ltd, Nanjing, China. He was a guest researcher with Ørsted Wind Power, Fredericia, Denmark, from November to December, 2018. His research interests include the modelling and stability analysis of the power electronic based power systems.

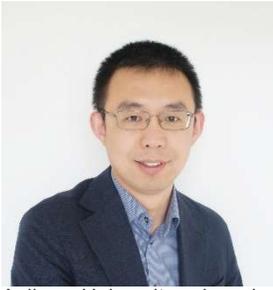

**Xiongfei Wang** (S'10-M'13-SM'17) received the B.S. degree from Yanshan University, Qinhuangdao, China, in 2006, the M.S. degree from Harbin Institute of Technology, Harbin, China, in 2008, both in electrical engineering, and the Ph.D. degree in energy technology from Aalborg University, Aalborg, Denmark, in 2013.

Since 2009, he has been with the Department of Energy Technology, Aalborg University, where he became Assistant Professor in 2014, an Associate Professor in 2016, a Professor and Research Program Leader for Electronic Power Grid (eGrid) in 2018. His current research interests include modeling and control of grid-interactive power converters, stability and power quality of power electronic based power systems, active and passive filters.

Dr. Wang was selected into Aalborg University Strategic Talent Management Program in 2016. He received six IEEE prize paper awards, the 2017 outstanding reviewer award of IEEE TRANSACTIONS ON POWER ELECTRONICS, the 2018 IEEE PELS Richard M. Bass Outstanding Young Power Electronics Engineer Award, and the 2019 IEEE PELS Sustainable Energy Systems Technical Achievement Award. He serves as an Associate Editor for the IEEE TRANSACTIONS ON POWER ELECTRONICS, the IEEE TRANSACTIONS ON INDUSTRY APPLICATIONS, and the IEEE JOURNAL OF EMERGING AND SELECTED TOPICS IN POWER ELECTRONICS.